\documentstyle[12pt]{article}

\begin{document}

\begin{center} {\large\bf A Phase in a Coherent State Wave Function --
Is It Always Irrelevant?}\\[3mm]
{\bf W.~Berej and P.~Rozmej}\\[2mm]
{\it 
Theoretical Physics Department, University Maria Curie-Sk\l odowska,
Pl.~M.Curie-Sk\l odowskiej 1, Pl-20031 Lublin, Poland\\
E-mail: berej@tytan.umcs.lublin.pl, rozmej@tytan.umcs.lublin.pl}\\
\today
\end{center}

\vspace{3mm}
The coherent states of a harmonic oscillator were first constructed
by Schr\"{o}dinger$^{1,2}$ as "the most classical" states of the 
oscillator.
The properties of these states have been studied in a systematic way
by Glauber$^{3}$, who showed their importance for the quantum 
mechanical 
treatment of optical coherence and who introduced the name "coherent 
state".
The description of the coherent states is now a standard subject 
in certain textbooks on quantum mechanics$^{4,5}$, monographs on 
quantum
optics$^{6,7,8}$ and more specialized texts devoted to coherent 
states of
different kinds$^{9,10,11}$. They were also discussed in many 
articles in
this journal$^{12}$. 

As the applications of the coherent states usually do not require the 
explicit
introduction of coordinate variable, the wave function for the 
coherent
state is written down only in some texts$^{3,5,8,9,10}$.
However, a phase factor, properly defined at the beginnig of the
discussion of the coherent states, is eventually omitted 
in some sources$^{3,8,10}$.
In this note we would like to point out the circumstance in which the 
knowledge
of this unimodular factor in the coordinate representation of the 
coherent
state is essential.

Let's recall first the basic facts concerning the coherent 
states$^{13}$.
Following Glauber$^{3}$, the coherent state 
$|\alpha\rangle$ is conveniently defined as an eigenstate of the
harmonic oscillator anihilation operator $\hat{a}$ with complex 
eigenvalue $\alpha$:
\begin{equation}
 \hat{a}\,|\alpha\rangle = \alpha\,|\alpha\rangle.
\end{equation}
Using this definition and its adjoint form, one can calculate the 
expectation 
values of the position and momentum
\begin{eqnarray}
\langle\alpha|x|\alpha\rangle &\!\! =\!\! 
     & (2\hbar/m\omega)^{\frac{1}{2}}\,\mbox{Re}\,\alpha\nonumber\\
\langle\alpha|p|\alpha\rangle & = 
     & (2\hbar m\omega)^{\frac{1}{2}}\,\mbox{Im}\,\alpha.
\end{eqnarray}
It is also not difficult to find the Fock representation of the 
normalized state
$|\alpha\rangle$ in the basis of states $|n\rangle$ labelled by the
occupation number $n$:
\begin{equation}
 |\alpha\rangle = e^{-\frac{1}{2}|\alpha|^{2}}\,\sum_{n=0}^{\infty}
                    \frac{\alpha^{n}}{\sqrt{n!}}|n\rangle,
\end{equation}
where the phase is defined by choosing
\begin{equation}
  \langle 0|\alpha\rangle= e^{-\frac{1}{2}|\alpha|^{2}}.
\end{equation}
In the further discussion of the coherent states one can check their 
well known
properties: they minimize the uncertainty relation and evolve in time,
following the motion of a classical oscillator. The effect of the 
time evolution
operator acting on the coherent state $|\alpha\rangle$ is easily 
calculated
by making use of expansion (3):
\begin{equation}
 \exp(-iHt/\hbar)|\alpha\rangle =
 e^{-\frac{1}{2}i\omega t}|\alpha e^{-i\omega t}\rangle,
\end{equation}
i.e. the coherent state remains coherent at all times. One can
immediately see that expectation values (2) of the position and 
momentum 
in the evolving coherent state carry out a simple harmonic motion.

Property (5) allows one to write down the wave function obeying 
the Schr\"{o}dinger equation and the initial condition 
$|\psi(0)\rangle=|\alpha\rangle$, simply adding the factor with the 
zero-point
energy in the exponent and substituting
$\alpha e^{-i\omega t}$ for $\alpha$ in the coordinate representation
of the coherent state $|\alpha\rangle$. However, to exploit this 
property
one must know the full coherent state wave function 
$\psi_{\alpha}(x)= \langle x|\alpha\rangle$ with a proper phase factor
defined by condition (4). As we shall see below, the phase
depends on the complex number $\alpha$ labelling the coherent state
and the substitution $\alpha\longrightarrow\alpha e^{-i\omega t}$ 
makes
it time dependent. Therefore, the omission of the phase factor 
leads to the wave function which is not a solution of 
the Schr\"{o}dinger equation. During our study of the dynamics 
of wave packets$^{14}$, among the cited references$^{3-12}$
we found only one text$^{5}$  which provides
this phase with a detailed derivation$^{15}$. 

To the end of this note we derive the needed phase factor starting 
from 
definition (1). It is a simple exercise involving calculation of 
the standard integrals with the Gaussian functions in their 
integrands, 
though certain care must be taken because $\alpha$ is a complex 
number.
At the same time we shall correct some errors existing in the 
literature.
The eigenvalue problem of Eq. (1) in the coordinate representation 
is a first-order differential equation for the wave function 
$\psi_{\alpha}(x)$.
Its general solution has the following form: 
\begin{equation}
\psi_{\alpha}(x) = 
  A\exp\!\left[-\left(\sqrt{\frac{m\omega}{2\hbar}}x-
\alpha\right)^{\!2\,}
       \right],
\end{equation}
where $A$ is a normalization constant. Thus, the corresponding 
probability 
density has a familiar, Gaussian shape.
The normalization condition yields$^{16}$
\begin{equation}
A = \left(\frac{m\omega}{\pi\hbar}\right)^{\!\frac{1}{4}} 
    \exp\left\{\frac{1}{4}(\alpha-\alpha^{\ast})^{2}+i\phi\right\},
\end{equation}
where $\phi$ is a real phase to be established from condition (4).
Carrying out one more integration, in which the explicit expression
for the ground state wave function of the harmonic oscillator is 
used, 
one obtains:
\begin{equation}
    i\phi = \frac{1}{4}(\alpha^{2}-\alpha^{\ast\,2}).
\end{equation}    
Therefore, the normalized coherent state wave function with a phase 
factor
chosen to ensure (4), may be written as:
\begin{equation}
\psi_{\alpha}(x) = 
\left(\frac{m\omega}{\pi\hbar}\right)^{\!\frac{1}{4}}
\exp\!\left(\frac{1}{2}\alpha^{2}-\frac{1}{2}|\alpha|^{2}\right)\,
 \exp\!\left[-\left(\sqrt{\frac{m\omega}{2\hbar}}x-
\alpha\right)^{\!2\,}\right].
\end{equation}
After the introduction of $\alpha = \alpha_{1} + i\alpha_{2}$, this 
may be
rewritten in a more readable form:
\begin{equation}
\psi_{\alpha}(x) = 
\left(\frac{m\omega}{\pi\hbar}\right)^{\!\frac{1}{4}}
\exp\left(-i\alpha_{1}\alpha_{2}\right)\,
\exp\!\left[-\left(\sqrt{\frac{m\omega}{2\hbar}}x-
\alpha_{1}\right)^{\!2\,}\right]\,
 \exp\!\left(2i\alpha_{2}\sqrt{\frac{m\omega}{2\hbar}}x\right).
\end{equation}
Finally, one can introduce the expectation values of Eq. (2), for 
brevity
denoted by $\langle x\rangle, \langle p\rangle$:
\begin{equation}
\psi_{\alpha}(x) = 
\left(\frac{m\omega}{\pi\hbar}\right)^{\!\frac{1}{4}}
\exp\!\left(-i\frac{\langle x\rangle\langle p\rangle}{2\hbar}\right)\,
 \exp\!\left[-\frac{m\omega}{2\hbar}(x-\langle x\rangle)^{2}\right]\,
 \exp\!\left(\frac{i}{\hbar}\langle p\rangle x\right).
\end{equation}
The above epressions may be found in some research papers$^{17}$.
We see that the coherent state wave function differs from a Gaussian
wave packet with the mean momentum $\langle p\rangle$ by the phase 
factor 
$\exp\!\left[-i\langle x\rangle\langle p\rangle/(2\hbar)\right]$
which must be included if the quantum evolution of $\psi_{\alpha}(x)$
is considered and Eq. (5) is used. When for special initial conditions
the substitution $\langle x\rangle\longrightarrow x_{0}\cos\omega t,
\langle p\rangle\longrightarrow -m\omega x_{0}\sin\omega t$ is made in
Eq. (11), one immediately obtains the well known wave packet 
oscillating 
without change of its shape$^{18}$:
\begin{equation}
\psi(x,t) = \left(\frac{m\omega}{\pi\hbar}\right)^{\!\frac{1}{4}}\
\!\exp\!\left[-\frac{m\omega}{2\hbar}(x\!-\!x_{0}\cos\omega 
t)^{2}\right]\,
\!\exp\!\left[-i\!\left(\frac{1}{2}\omega t\! +\!
        \frac{m\omega}{\hbar}x_{0}x\sin\omega t
        \!-\!\frac{m\omega}{4\hbar}x_{0}^{2}\sin2\omega 
t\right)\right].
\end{equation}

\vspace{8mm}
\noindent
{\bf Literature}
\small
\begin{enumerate}
\item E. Schr\"{o}dinger, "Der Stetige \"{U}bergang von der Mikro- 
zur 
Makromechanik", Naturwissenschaften {\bf 14}, 664--666 (1926).
\item See the interesting historical paper: 
F. Steiner, "Schr\"{o}dinger's Discovery of Coherent States",
Physica B {\bf 151}, 323--326 (1988).
\item R.J.~Glauber, "Coherent and Incoherent States of the Radiation 
Field",
Phys. Rev. {\bf 131}, 2766--2788 (1963).
\item E. Merzbacher, {\em Quantum Mechanics} (Wiley, New York, 1970),
second edition, pp. 362--370.
\item C. Cohen-Tannoudji, B. Diu, and F. Laloe, {\em Quantum 
Mechanics},
Vol. 1, (Wiley, New York, 1978), Complement G$_{V}$.
\item J.R. Klauder and E. Sudarshan, {\em Fundamentals of Quantum 
Optics}, 
(Benjamin, New York, 1968), Chap. 7.
\item W.H. Louisell, {\em Quantum Statistical Properties of Radiation}
(Wiley, New York, 1973), pp. 104--110.
\item L.~Mandel and E.~Wolf, {\em Optical Coherence and Quantum 
Optics} 
(Cambridge University Press, Cambridge, 1995), Chap. 11.
\item J.R. Klauder and B.-S. Skagerstam, {\em Coherent States. 
Applications
in Physics and Mathematical Physics} (World Scientific, Singapore, 
1985).
\item A.M. Perelomov, {\em Generalized Coherent States and Their
Applications} (Springer, Berlin, 1986).
\item W.-M. Zhang, D.H. Feng, and R. Gilmore, "Coherent States: 
Theory and
Some Applications", Rev. Mod. Phys. {\bf 62}, 867--927 (1990).
\item We mention only two of them: P. Carruthers and M.M. Nieto, 
"Coherent States and the Forced Harmonic
Oscillator", Am. J. Phys. {\bf 33}, 537--544 (1965);
S. Howard and S.K. Roy, "Coherent States of a Harmonic Oscillator",
{\em ibid.} {\bf 55},1109--1117 (1987).
\item Sections III and IV of Glauber's paper (Ref. 3) contain
an excellent discussion of the properties of the coherent states and 
should 
be an obligatory reading for anyone interested in this subject.
A complete and thorough discussion may also be found in Ref. 5. 
The authors of this textbook choose as the defining condition 
of the coherent state the requirement that 
the mean values $\langle x\rangle, \langle p\rangle, \langle 
H\rangle$ 
are as close as possible to the classical values. 
\item R.~Arvieu, P.~Rozmej, and W.~Berej, "Time-dependent partial 
waves
and vortex rings in the dynamics of wavepackets", J. Phys. A {\bf 30},
5381--5392 (1997) .
\item The correct phase factor is also given in Ref. 7. However, the 
order
of the reasoning is reversed there: the phase is chosen somewhat
arbitrarily and then $\langle x|\alpha\rangle$ serves as the 
generating 
function for oscillator eigenfunctions.
\item The reader of Ref. 3 should be warned that the coherent state 
wave funcions given by the formula (3.29) of this paper are actually 
not normalized: 
the exponential present in our expression (7) is missing there. 
In the normalization constant written in Ref. 8 there is a sign error
in the exponent.
\item For example: M.M. Nieto and L.M. Simmons Jr., "Coherent States 
for General
Potentials I", Phys. Rev. D {\bf 20}, 1321--1331 (1979); S.-J. Chang 
and
K.-J. Shi, "Evolution and Exact Eigenstates of a Resonant Quantum 
System",
Phys. Rev. A {\bf 34}, 7--22 (1986).
\item L.I.~Schiff, {\em Quantum Mechanics} (Mc-Graw Hill, New York, 
1968),
third edition, pp. 74--76.

\end{enumerate}

\end{document}